\documentclass[preprint, pra, amsmath, amssymb, aps, groupedaddress]{revtex4-2}
\usepackage{physics}
\usepackage{graphicx}
\usepackage[dvipsnames]{xcolor}
\newcommand{\keff}{k_\mathrm{eff}}
\newcommand{\keffvec}{\mathbf{k}_\mathrm{eff}}

\begin{document}

%\preprint{APS/123-QED}

\title{Real-time phase control methods for cold-atom interferometry}

\author{Mohamed Guessoum}
\author{Nathan Marlière}
\author{Charbel Cherfan}
\author{Remi Geiger}
\author{Arnaud Landragin}
    \email{arnaud.landragin@obspm.fr}

\affiliation{LTE, Observatoire de Paris, Université PSL, Sorbonne Université, Université de Lille, LNE, CNRS, 61 avenue de l’Observatoire, 75014 Paris, France}

\date{\today}

\begin{abstract}

We present two methods to achieve real-time inertial phase compensation in atom interferometers. Both methods, based on jumps of the position of the retroreflection mirror or frequencies of Raman lasers, demonstrate similar state-of-the-art performance on our cold atom gyroscope, comparable to that of the reference method based on optical phase jumps. These alternative approaches broaden the scope of applications for real-time inertial phase compensation methods in atomic interferometers, particularly for space applications.
\end{abstract}

\maketitle

\section{Introduction}

\color{Black}
Inertial sensors based on cold atom interferometry have demonstrated both high sensitivity and accuracy~\cite{geigerHighaccuracyInertialMeasurements2020} for applications in inertial guidance, geophysics or test of fundamental physics. In most of these sensors, the sensitivity is limited by the sequential nature of the measurements leading to the sampling of acceleration and rotation signals, and by the dead times between measurements.  Various methods have been developed to reduce this source of noise and its impact on performance, such as the use of an isolation platform to reduce its contribution~\cite{hensleyActiveLowFrequency1999} and the correlation with conventional sensors to estimate the inertial noise~\cite{merletOperatingAtomInterferometer2009}. A further step was to hybridize the quantum and conventional sensors~\cite{lautierHybridizingMatterwaveClassical2014} in order to benefit from their respective advantages of bandwidth and long-term stability. This includes aspects of sensor data fusion on the one hand, and real-time phase compensation (RTC) of vibration noise in the atomic interferometer on the other. In particular, RTC ensures that the interferometer operates at its optimum sensitivity in the middle of the interferometer's central fringe. The use of these methods of hybridization has made it possible to extend the field of application to transportable terrestrial gravimeters~\cite{menoretGravityMeasurements1092018}, on-board gravity measurements~\cite{bidelAbsoluteMarineGravimetry2018,jensenAirborneGravimetryQuantum2024}, possibly gravity and acceleration strap-down measurements~\cite{templierTrackingVectorAcceleration2022} or acceleration and rotation measurements simultaneously~\cite{darmagnacdecastanetAtomInterferometryArbitrary2024,salducciQuantumSensingAcceleration2024}. 

RTC of the atomic vibration phase alone has been shown to optimize the signal-to-noise ratio of a quantum sensor for gravity measurements~\cite{lautierHybridizingMatterwaveClassical2014, menoretGravityMeasurements1092018} or acceleration measurements~\cite{darmagnacdecastanetAtomInterferometryArbitrary2024} but also for rotation measurements~\cite{duttaContinuousColdAtomInertial2016,luRealtimeCompensationScheme2025}. All these sensors are based on the use of two-photon Raman transitions to manipulate atomic wave packets. The method involves applying inertial phase compensation via a phase jump on one of the lasers just before the last recombination pulse. Unfortunately, if we consider atomic sensors in weightlessness, where the two photon transitions are based on the retroreflection of the Raman~\cite{barrettDualMatterwaveInertial2016,heSpaceColdAtom2023} or Bragg~\cite{lachmannUltracoldAtomInterferometry2021} lasers, the phase difference between the two lasers no longer plays any role~\cite{levequeEnhancingAreaRaman2009}. This is of particular importance for space missions proposed to test fundamental physics~\cite{aguileraSTEQUESTtestUniversalityFree2014} or geodesy~\cite{levequeGravityFieldMapping2021}. This work demonstrates alternative methods to achieve real-time control compatible with space or weightlessness sensors and, more generally, sensors that do not use Doppler shift to lift the degeneracy from opposite Raman or Bragg diffraction processes in retroreflected configuration.

The article first presents our atomic gyroscope setup, then successively the two new methods based on mirror jumps and frequency jumps, and finally the comparison of the sensitivity to rotation measurement for all the methods. The relevance of these alternative methods is demonstrated by comparing their effectiveness with that of the reference method using phase jumps between Raman laser beams.

\section{Experimental setup}

\begin{figure}
    \centering
    \includegraphics[width=1\linewidth]{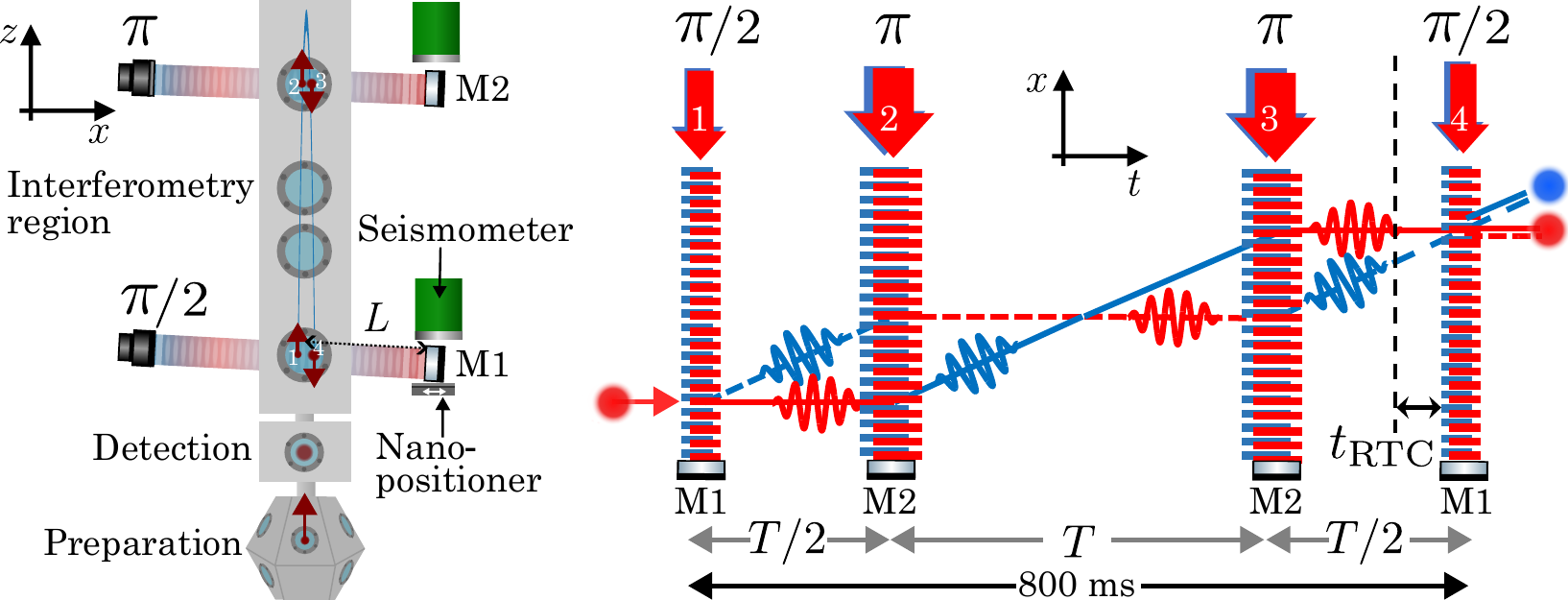}
    \caption{(Left) Scheme of the experimental setup. Atomic clouds are prepared at the bottom of the chamber, launched vertically and interact with two pairs of Raman lasers in a sequence of four pulses, numbered from 1 to 4, before falling back into the detection area. M1 and M2: retroreflection Raman lasers mirrors. L: distance from the mirror M1 to the positions of the atom cloud at the fourth pulse. The Raman laser beams are slightly tilted by 4° in the vertical direction. (Right) Diagram of the space-time diagram of the interferometer sequence. Beamsplitters and mirrors are realized by a set of Raman pulses of areas $\pi/2$ and $\pi$. The real-time compensation of the vibrations is applied just at a time $t_{RTC}$ before the beginning of the last beamsplitter pulse. The laser beams at frequency $\omega_{3}$  (respectively $\omega_{4}$) are represented in red (respectively blue). $\pi/2$ pulses (respectively $\pi$ pulses) are retroflected on mirror M$_1$ (respectively M$_2$)}
    \label{Fig:interfero}
    
\end{figure}

The experimental setup is described in previous work~\cite{duttaContinuousColdAtomInertial2016,savoieInterleavedAtomInterferometry2018}. 
Cesium atoms are trapped and cooled inside a 3D magneto-optical trap (MOT) loaded from an adjacent 2D MOT. Using the moving molasses technique, the atoms are then launched upwards, towards the interferometry region, and selected in state \(\ket{F=4,m_{F}=0}\). They then undergo an interferometric sequence with a total duration of $2T=$ 800~ms comprising four pulses, with area $\pi/2-\pi-\pi-\pi/2$, separated by time intervals $T/2-T-T/2$ as sketched in Fig.~\ref{Fig:interfero}. At the end of the sequence, the phase shift of this state-labeled interferometer is extracted by measuring the transition probability to the state \(\ket{F=3,m_{F}=0}\) by fluorescence means. We apply the phase compensation before the fourth pulse, when the evaluation of the vibration impact can be carried out for almost the entire duration of the interrogation. Nevertheless, a time delay of $t_\text{RTC}$ is required for it to be fully effective at the time of the last pulse. The phase jump allows for compensation of the inertial phase and places then the measurement point at the mid-fringe, where the phase sensitivity is maximal.

Atomic beamsplitter ($\pi/2$) and mirror ($\pi$) pulses are achieved by using two photon Raman transitions between the $\ket{F=3}$ and $\ket{F=4}$ hyperfine levels of the ground state. Let's call L3 and L4 the pair of Raman lasers, tuned close to the D2 line at 852~nm. The frequency $\omega_{3}$ of L3 (respectively $\omega_{4}$ of L4) is red detuned by $\Delta$ to the \(\ket{F=3}\) relative to \(\ket{F'=3}\) transition (respectively $\Delta$ to the \(\ket{F=4}\) to \(\ket{F'=3}\) transition). Lasers L3 and L4 are phase-locked and have a frequency difference  $\delta\omega_L=(\omega_{3}-\omega_{4})$ in the vicinity of hyperfine splitting, with both phase and frequency adjustable in real time. The associated wavelengh of laser L3 (respectively L4) is $k_3=\omega_3/c$ (respectively $k_4=\omega_4/c$). 

In such an interferometer, when the hamiltonian is at most quadratic in position and momentum, the total phase shift between the two arms can be expressed as a function of the phase difference between the Raman lasers at the center of the atomic wave-packets at the pulse times~\cite{bordeAtomicClocksInertial2002}. It can be written in the referential frame of the apparatus as follows:
\begin{equation}
\Delta\Phi=\Phi_{Inert}+\left[\Phi(0)-2\Phi\left(\dfrac{T}{2}\right)+2\Phi\left(\dfrac{3T}{2}\right)-\Phi(2T)\right],  \label{eq:phase}
\end{equation}
where $\Phi_{Inert}$ is the accumulated phase related to the displacement of the atoms due to inertial forces, and $\Phi(t)$ is the difference between the optical phases $\phi_3$ and $\phi_4$ of Raman lasers at the positions of the unperturbed trajectory $\mathbf{r}$.  This second term in Eq.\ref{eq:phase} cancels out in the case of perfect, undisturbed Raman lasers.

In practice, we use a configuration in which the two Raman lasers propagate together and are retroreflected on mirrors. We can then apply the so-called k-reversal technique to reduce the contribution of some of the systematic effects~\cite{weissPrecisionMeasurementMCs1994}. Atoms can then be selectively diffracted in one direction or in the opposite direction, with a transfer of momentum $\pm\hbar \mathbf{k_{eff}}$, where $\mathbf{k_{eff}}=\mathbf{k_3}-\mathbf{k_4}$. To enable this, the direction of the retroreflecting mirrors is slightly tilted by 4° with respect to the horizontal direction in order to lift the degeneracy of resonance conditions due to the opposing Doppler shifts caused by gravity. In this configuration, only one of two retroreflected Raman laser pairs interacts with the atoms. Moreover, the Raman lasers do not accumulate relative phase noise between the point where they are locked together and the entrance to the vacuum chamber where the atoms are manipulated~\cite{duttaContinuousColdAtomInertial2016}. The phase difference between the two lasers driving the transition is then given by: 
 \begin{flalign}
        \Phi^{\pm}(t)  &= \left(\omega_3 t + \phi_3^0 -\mathbf{k}_3^{\pm}\cdot \mathbf{r}\right) - \left( \omega_4 t + \phi_4^0 - \mathbf{k}_4^{\pm}\cdot \mathbf{r} \right) \notag \\
                       &=  \delta\omega_L t + \phi_L \mp\keffvec \cdot \mathbf{r},  \label{eq:plus-minus_phase}
 \end{flalign}
 
 where the subscripts + and - indicate the direction of the Raman transition selected to drive the transition, $\phi_L=\phi_3^0-\phi_4^0$ the difference of absolute phase of the two Raman lasers at the entrance of the vacuum chamber, and $\mathbf{r}$ is defined in the referential frame of the instrument. 
   
\section{REAL-TIME COMPENSATION METHODS}
In our interferometer, the phase due to the difference in laser frequencies cancels out over the four pulses when it is constant (as well as in a three-pulse interferometer case) or linearly ramped. The remaining laser terms to be taken into account for the rest of this paper are as follows:

 \begin{flalign}
        \Phi^{\pm}(t) = \phi_L \mp\keffvec \cdot \mathbf{r}.     \label{eq:plus-minus_phase-final}
 \end{flalign}
                       
\begin{figure}
    \centering
    \includegraphics[width=1\linewidth]{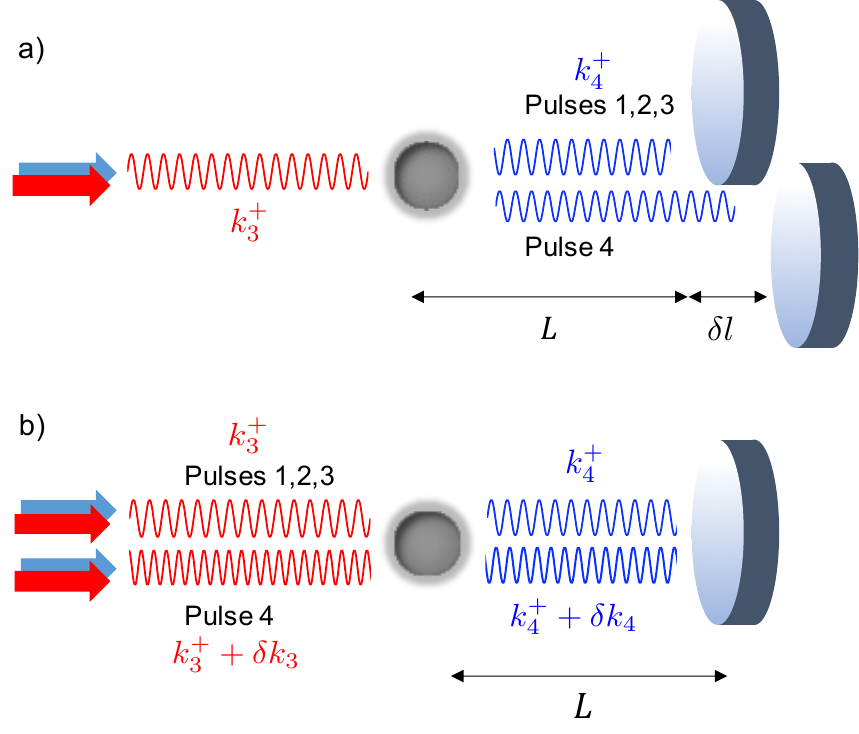}
    \caption{Schematic diagram of the two alternative methods developed in order to control the output phase of the interferometer. The two Raman lasers are brought together in the experiment and retroreflected on the mirror. Real-time compensation is achieved by modifying the optical phase difference between the two counter-propagating Raman light fields at the position of the atoms before the last pulse: a) by moving the retroreflecting mirror and b) by modifying the effective wave vector of the Raman transition. Only the laser fields involved in the transition are shown.}
    \label{fig:saut}
\end{figure}

So far, RTC of inertial phase by laser phase has been demonstrated by adjusting the phase difference $\phi_L$ just before the last pulse~\cite{lautierHybridizingMatterwaveClassical2014}. This method compensates $\Phi_{Inert}$ from an estimate of classical sensor measurements made during the interrogation. It has been shown to implement the compensation phase well enough to become negligible compared to the estimation error itself. It will be used as a reference method to verify the effectiveness of the other two methods and is detailed in the reference~\cite{savoieInterleavedAtomInterferometry2018}. 
But, we can see from the second term of Eq.\ref{eq:plus-minus_phase-final} that two other parameters can be used to change the optical phase and achieve this compensation: the position of the retroreflecting mirror relative to the atom's trajectory and the modulus of the effective wave vector. These two methods are sketched in Fig.~\ref{fig:saut} and will be described in detail in the following sections.

 \subsection{Mirror position jumps}
    The first approach to control the phase of the interferometer consists of changing the position of one of the retroreflecting mirrors. As the first and last pulses (respectively second and third pulses) are realized by retroreflection on the same mirror, it's absolute distance $L$ to the atoms do not play any role. But a displacement of $\delta l$ of the reflective mirror before the last pulse leads to an increment to the total phase shift by: 
    
 \begin{equation}
        \Delta\Phi^{\pm}=\mp \keff \, \delta l.
        \label{eq:mirror_jump}
    \end{equation}
     
\begin{figure}
    \centering
    \includegraphics[width=1\linewidth]{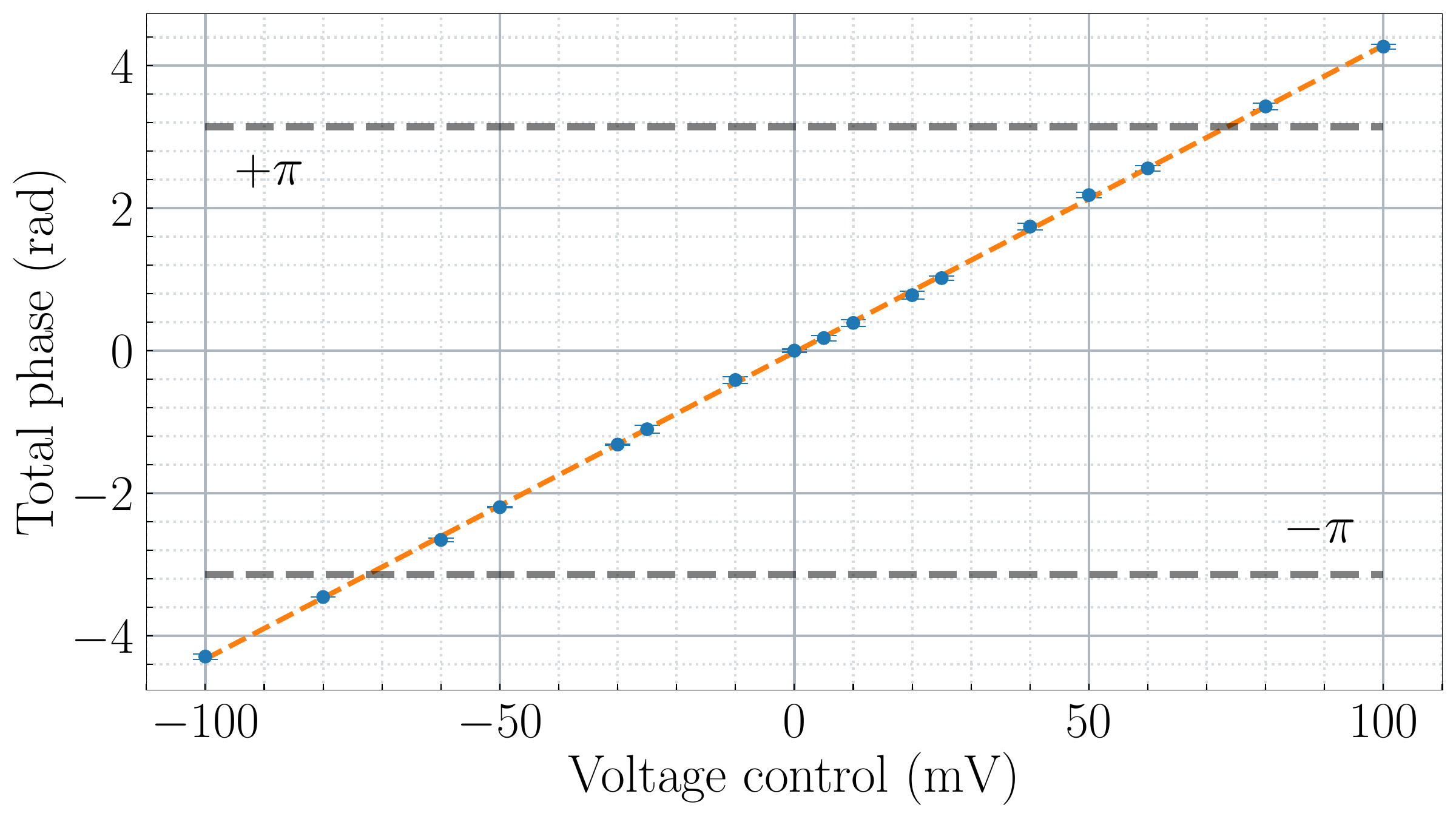}
    \caption{Variation of the interferometer phase as a function of the mirror displacement between the third and fourth pulses. The x-axis shows the control voltage of the nanopositioner. Dots are the experimental results average over 700 cycles. Dashed line is the linear fit with a slop of  $43.08(15) \,\mathrm{rad.V}^{-1}$.}
    \label{Fig:Calibration-position}
\end{figure}

    Experimentally, this is achieved by mounting the retroreflecting mirror M$_1$ on a voltage controlled nanopositioner (Mad City Labs model Nano-OP30M). Spanning one period of the fringes can be done by moving the mirror by a distance $\delta l = \lambda/2$, i.e. $\sim 426.2\,\mathrm{nm}$. The mechanical displacements follow an nearly exponential motion with a time constant of 7.5~ms. Phase compensation is implemented with $t_{RTC}= 40\ \mathrm{ms}$, so that the position change is established to better than 99$\%$ and $t_{RTC}$ is a multiple of $1/50\ \mathrm{Hz}$ (the frequency of the electrical network that causes noise on seismometer signal acquisition). The exact scale factor of the nanopositioner was determined by integrating measurements of the total interferometer phase while performing constant mirror position jumps before the last $\pi/2$-pulse for different displacement values (Fig.~\ref{Fig:Calibration-position}). The characterization results in a scale factor $43.08(15) \, \mathrm{rad.V}^{-1}$ constant over the whole range of applied control voltage. The result correspond to a displacement of the mirror of  $2.92(1)\ \mu \mathrm{m.V^{-1}}$ in agreement with the manufacturer specification of $3.01\ \mu \mathrm{m.V^{-1}}$.

 \subsection{Frequency jumps}

    The second method consists in modifying the effective wave vector of the Raman transition. This is achieved by changing the Raman detuning by a quantity $\delta\Delta$, that means the frequency of both Raman lasers, $\delta\omega_3=\delta\omega_4=\delta\Delta$, while keeping them in phase with each other at a constant frequency difference. When applied before the last pulse, the phase shift added to the interferometer is:
    
   \begin{flalign}
         \Delta\Phi^{\pm} &=\mp L \, \delta\keff
                = \mp\delta\Delta \, \tau,
        \label{eq:freq_jump}
    \end{flalign}

    where $\delta \keff=2\keff.\delta\Delta/(\omega_3+\omega_4)$ is the change of the wave vector and $\tau = 2L/c$ the delay of retroreflection from the atom to the mirror and back. This sensitivity of the interferometer phase to the change of Raman laser frequency has been identified and characterized as a potential source of noise~\cite{mcguirkSensitiveAbsolutegravityGradiometry2002,legouetInfluenceLasersPropagation2007} but also a way for compensating for~\cite{rouraCircumventingHeisenbergsUncertainty2017,damicoCancelingGravityGradient2017} and measuring~\cite{caldaniSimultaneousAccurateDetermination2019,janvierCompactDifferentialGravimeter2022} the gradient of gravity. In our experiment, as the distance $L$ is approximately $10 \,\mathrm{cm}$, we needed a frequency jump of about $\pm 800\, \mathrm{MHz}$ in order to achieve a $\pm \pi$ phase shift. However, changing the frequencies of the Raman laser requires further modifications to the laser bench in order to keep first both lasers phase locked and second the coupling with the atoms, characterized by the Rabi frequency, constant. 
    
\begin{figure}
        \centering
        \includegraphics[width=1\linewidth]{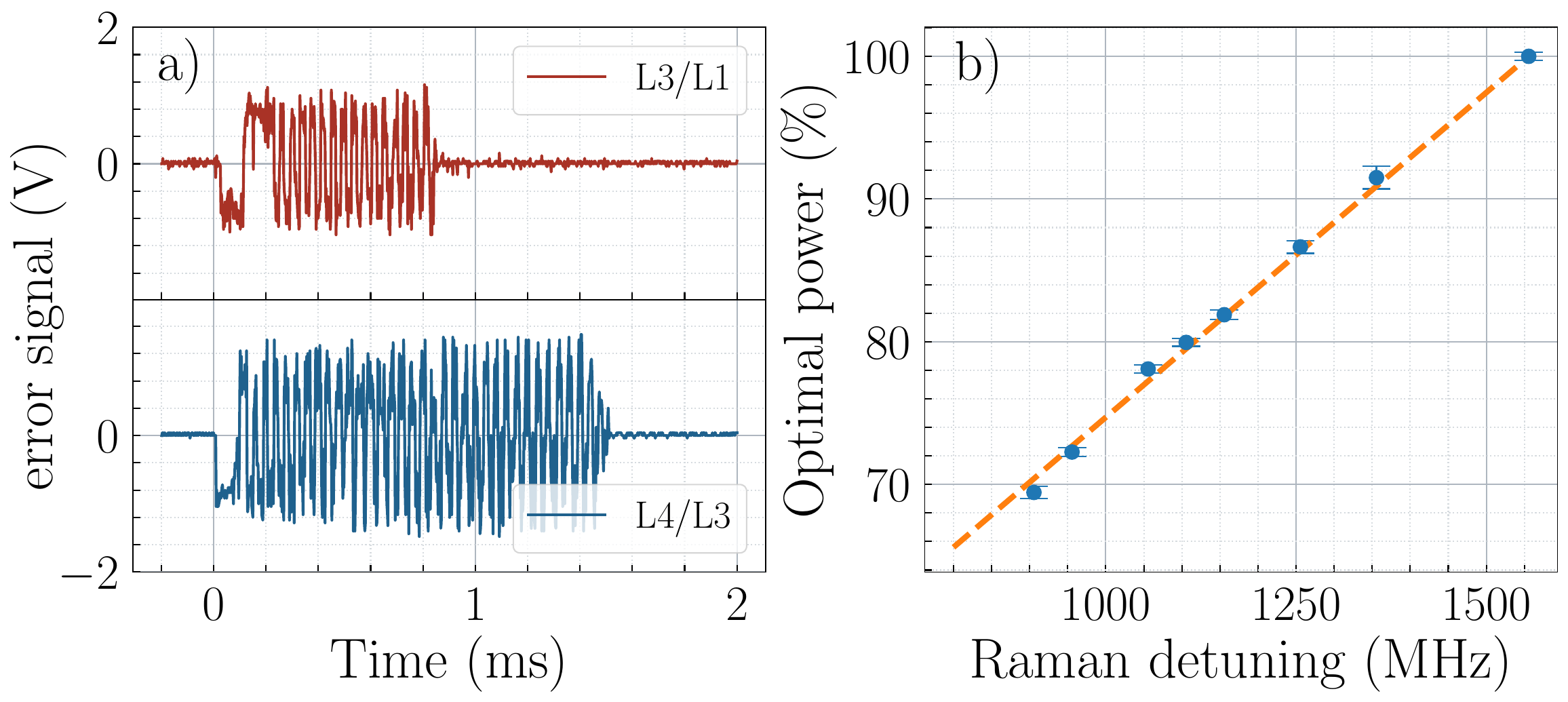}
        % \noindent
        % \resizebox{1.2\linewidth}{!}{\input{Figure4.pgf}}
        \caption{Characterization of modifications to the Raman bench in order to achieve a frequency jump while keeping the Rabi frequency constant: a) post-frequency jump time evolution of error signals from the frequency and phase control system of laser L3,  on the top, and L4, on the bottom, versus reference-locked to laser L1 and laser L3 respectively (see text). b) relative optimum power to preserve the $\pi$ or $\pi/2$ pulse conditions as a function of Raman detuning. The reference point is set at maximum Raman detuning for $\abs{\Delta}=1555\ \mathrm{MHz}$. Each point is deduced from the Rabi oscillations based on measurement over 30 cycles. The Dashed orange line is a linear fit of the data points with a slope of $4.6$\% variation in optimum power per $100\ \mathrm{MHz}$ variation of the Raman detuning.}
        \label{fig:pulse-adjustment}
    \end{figure}

    The two lasers L3 and L4 are extended-cavity lasers that can be phase and frequency-stabilized by a piezoelectric actuator and diode current before being optically amplified. Each error signal is generated from a beatnote with a master laser, which is compared to a signal controlled by an RF direct digital synthesizer (DDS).  The laser L3 is frequency locked with respect to the repumping laser L1, allowing adjustment of the detuning of the Raman transition from $\Delta=-755\, \mathrm{MHz}$ to $\Delta=-1555\, \mathrm{MHz}$. The L4 Laser is then locked to the laser L3 and follows it when the latter is modified. The frequency range is chosen around $\Delta=-1155\, \mathrm{MHz}$ where the dependence of the Raman laser light shift on detuning is minimal for Cesium atom~\cite{weissPrecisionMeasurementMCs1994}. The light shift stays below $12$\%  of the Rabi frequency, which allows to keep the maximum contrast of the interferometer. To prevent mode hopping during large frequency jumps, a synchronized feedforward mechanism is implemented, simultaneously adjusting both the diode current and the piezoelectric actuator voltage of each laser. Nevertheless, the frequency jump is limited to a range of 800~MHz total in order to avoid any unlocking event. The laser L3 (respectively L4) is ready for performing the Raman pulse in less than 1~ms (respectively 1.6~ms), see Fig.~\ref{fig:pulse-adjustment}(a). The jump of frequency is done with $t_{RTC}= 20\ \mathrm{ms}$, so that the frequency change is established and $t_{RTC}$ is a multiple of $1/50\ \mathrm{Hz}$. 
    
    The second modification to the Raman laser bench is to allow the optical power to be adjusted to compensate for the change in coupling of the Raman transition, i.e. the Rabi frequency, which scales almost as $1/\Delta$, and to preserve the $\pi/2$ and $\pi$ pulse conditions. This adjustment is carried out on both beams simultaneously, using an acousto-optic modulator already in use for pulse shaping. All changes to the Raman lasers associated with frequency jumps during the sequence were calibrated in advance (see Fig.~\ref{fig:pulse-adjustment}(b) for the calibration of the laser power compensation) and fed back into the computer controlling the experiment. 

\begin{figure}
    \centering
    \includegraphics[width=0.95\linewidth]{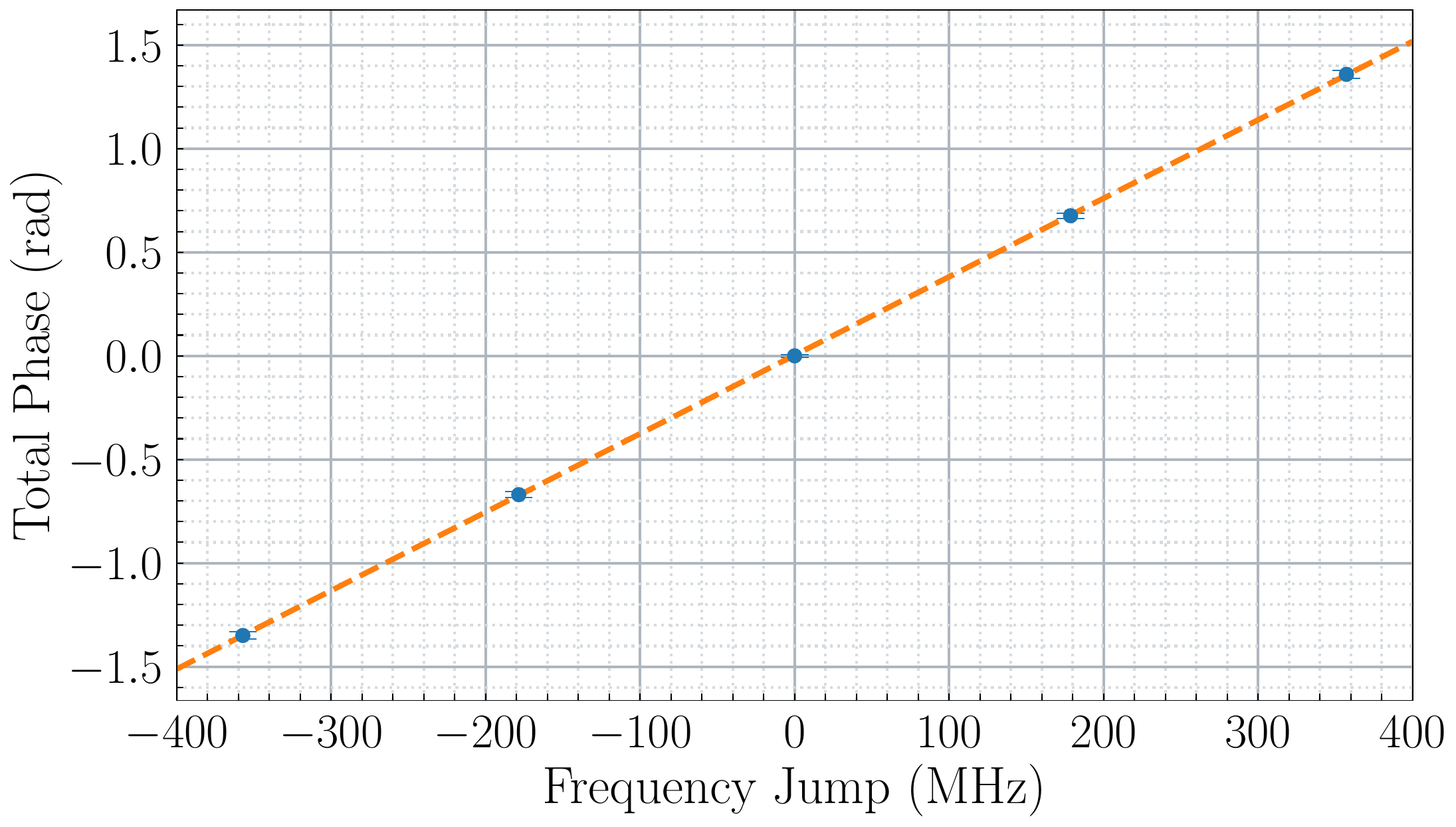}
    \caption{Atom interferometer phase shift as a function of the frequency jump before the last pulse, while keeping $\Omega_\text{Rabi}$ constant. Dots are the experimental results average over 700 cycles. The dashed line represents a linear fit of the experimental data with a slope of the $3.786(6)\ \mathrm{mrad/MHz}$.}
    \label{fig:Calibration-frequency}
\end{figure}

    We determine experimentally the scaling factor of this method. Indeed, if the wave vector is very well known, the distance $L$ of the atoms to the reflecting mirror has to be determined by a direct measurement with the interferometer~\cite{biedermannTestingGravityColdatom2015}. Fig.\ref{fig:Calibration-frequency} shows the determination of the change of the output phase of the interferometer for a variation of the detuning around $\Delta=-1155\ \mathrm{MHz}$. The fit leads to an average distance of $L=90.4(2)\ \mathrm{mm}$. The value of the distance $L$, combined with the frequency tuning range of the lasers, enables the interferometer phase to be adjusted over a full range of $\pi$. In order to cover the entire range of 2$\pi$ required for RTC methods, the distance L could be doubled, as could the frequency jump range, which requires a change in Raman lasers to prevent unlocking. This full range condition would also be achieved in a double diffraction configuration in which the induced atomic phase jump is twice as large.
    
\section{PERFORMANCES OF REAL-TIME PHASE COMPENSATION METHODS}

The performance comparison of real-time inertial phase compensation methods is carried out on our gyroscope as described in~\cite{savoieInterleavedAtomInterferometry2018} but in a continuous measurement configuration that is not interleaved, as in~\cite{duttaContinuousColdAtomInertial2016}. The inertial vibration phase is estimated from broadband seismometers by integrating the impact of the inertial signal weighted by the interferometer sensitivity function~\cite{lautierHybridizingMatterwaveClassical2014}. The real-time compensation is applied with a $t_{RTC}$ set at 20~ms for phase and frequency jumps and 40~ms for position jumps. In order to reduce the error due to the absence of signals for the last duration $t_{RTC}$, we extrapolate the estimate to the end of the sequence. In practice we used the signal recorded during a duration $t_{RTC}$ before the application of the compensation as a good estimation of the signal after it, as the bandwidth of the seismometer is limited to 50 Hz. In the case of position jumps, the compensation phase is integrated over successive measurements and converted modulo $2\pi$ into mirror position, remaining within a range of $\lambda/2$ around the initial position and and avoiding cumulative position errors. In the case of frequency jump compensation, this is limited to a phase shift of $\pi$. This is why only half of the compensation of the inertial phase is applied via a frequency jump, while the remaining half is implemented through a phase jump. In the same way, we have applied the total compensation calculated modulo $2\pi$ so that the frequency remains around the initial frequency for $\Delta=-1155\ \mathrm{MHz}$. The atomic phase signal is acquired every cycling time, i.e. 800~ms, and is converted in rotation rate measurement using the scaling factor of the gyroscope~\cite{gautierAccurateMeasurementSagnac2022} of $4.62\cross 10^{6}\ \mathrm{rad.s^{-1}/rad}$. 

\begin{figure}
    \centering
    \includegraphics[width=1\linewidth]{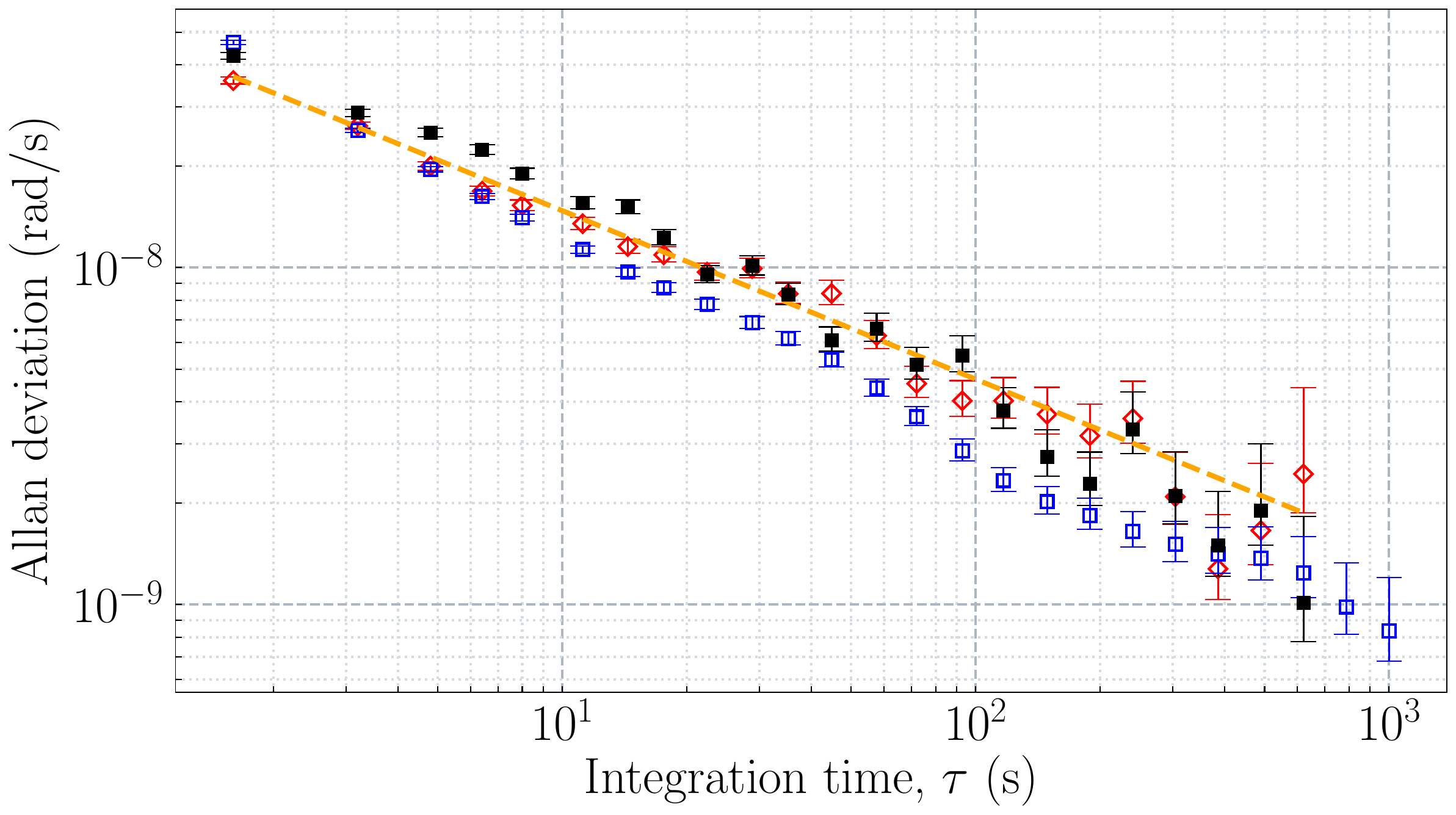}
    \caption{Standard Allan deviations of rotation measurement stability with different real-time compensation methods: in black, the reference method by the jumps of the difference in Raman laser phases, in blue by mirror position jumps and in red by combined frequency and phase jumps. The dashed line represents a decay in $1/\sqrt{\tau}$ characteristic of a white noise.}
    \label{fig:ADEV}
\end{figure}

Fig.~\ref{fig:ADEV} shows the comparison of Allan standard deviations (ADEV) of gyroscope stability with the three RTC methods. Similar behaviors are observed for all three methods, with ADEV decreasing as $\tau^{-1/2}$, illustrated by the dashed line. The gain in stability using RTC methods is estimated at a factor of 7 by comparing the ADEV of the estimated vibration phase (unplotted) with that of the interferometer phase measurements after compensation, which confirms the effectiveness of all methods. In addition, the sensitivity of around $5.10^{-8}\ \mathrm{rad.s^{-1}}$ at 1~s is similar to that of the state of the art~\cite{savoieInterleavedAtomInterferometry2018} if we consider that this experiment was not performed in interleaved mode.

\section{DISCUSSION and CONCLUSION}

We have demonstrated two new methods in order to realize real-time compensation for parasitic vibrations that's allows for state of the art performances in our cold atom gyroscope similar to these of the reference method based on phase jump between the two Raman lasers. 

The first method based on mirror position jump is the easiest to implement. There are, however, three points to consider. Firstly, the minimum time required for the position change to be complete is about 30 ms, which results in a delay that can prevent an accurate estimation of the vibrational phase. Secondly, special care must be taken to avoid a systematic error from cumulative position displacement if the mean jump is not zero, due to uncertainty in the knowledge of the nanopositioner scaling factor. In our case, in this preliminary study of the scaling factor with a relative uncertainty of $3.10^{-3}$, we can already guarantee that the bias on the rotation measurement corresponds to less than $10^{-9}$~rad.s$^{-1}$. Thirdly, the method is based on physical displacement, which can lead to synchronous vibrations and possibly to a source of long-term drift or bias. These three points do not represent a limitation in our gyroscope and can be specifically addressed and reduced in additional studies for each of the applications envisaged.

The second method, based on frequency jumps, has the advantage of using only opto-electronic means, with no moving mechanical parts. The jumps can be almost instantaneous, especially when they are the result of direct frequency synthesis, enabling the compensation phase to be calculated right up to the last moment before the pulse. Furthermore, the opto-mechanical method for implementing this compensation is the same as for gradient compensation and can therefore be added almost for free. A point of concern is the need to know the distance from the atomic cloud to the retroreflective mirror. As it can be calibrated directly, this is not a limitation in most cases where this distance is stable over time, but it may need to be addressed specifically for application onboard a moving vehicle in which acceleration will lead to non linearity in the compensation process. Finally, this method is slightly more complex, as it requires correction of the Rabi frequency change by adjusting the laser power. Another improvement would be to independently adjust the power of the two Raman laser beams, and not just their sum, in order to ensure compensation for light shift over a wider detuning range.

In conclusion, these two methods are complementary to the reference method based on phase jump, and can be used more generally for both Raman and Bragg diffraction, and in particular when using double-diffraction processes~\cite{levequeEnhancingAreaRaman2009,gieseDoubleBraggDiffraction2013}. Modifications to the gyroscope are currently being prepared in order to use the double diffraction method, which will enable the study of the contribution of its new RTC methods in this specific context. Last but not least, the applications for which these methods are most interesting aim to achieve the best performance in the absence of the Doppler effect, as in space equipments~\cite{aguileraSTEQUESTtestUniversalityFree2014,levequeGravityFieldMapping2021,heSpaceColdAtom2023}, or on the ground for horizontal beam splitters~\cite{canuelExploringGravityMIGA2018}, or without the possibility of changing the relative phase between laser beams.

\section{ACKNOWLEDGMENTS}
We acknowledge the financial support from Centre National d’Etudes Spatiales (CNES), from Agence Nationale pour la Recherche (project PIMAI, ANR-18-CE47-0002-01) and from a government grant managed by the Agence Nationale de la Recherche under the Plan France 2030 with the reference ”ANR-22-PETQ-0005”. M.G. acknowledges support from Region Ile-de-France, DIM SIRTEQ. The authors wish to thank L. Sidorenkov and R. Duverger for their careful reading. 

\color{Black}

\bibliography{RTC-mirror-frequency.bib}

\end{document}